# SmartLLMSentry: A Comprehensive LLM Based Smart Contract Vulnerability Detection Framework

Oualid ZAAZAA
*Mohammed V University in Rabat*
*Morocco*
oualid_zaazaa@um5.ac.ma
0000-0003-4864-2486
(Corresponding Author)

Hanan EL BAKKALİ
*Mohammed V University in Rabat*
*Morocco*
hanan.elbakkali@um5.ac.ma
0000-0003-2941-3768

*Abstract*— Smart contracts are essential for managing digital assets in blockchain networks, highlighting the need for effective security measures. This paper introduces SmartLLMSentry, a novel framework that leverages large language models (LLMs), specifically ChatGPT with in-context training, to advance smart contract vulnerability detection. Traditional rule-based frameworks have limitations in integrating new detection rules efficiently. In contrast, SmartLLMSentry utilizes LLMs to streamline this process. We created a specialized dataset of five randomly selected vulnerabilities for model training and evaluation. Our results show an exact match accuracy of 91.1% with sufficient data, although GPT-4 demonstrated reduced performance compared to GPT-3 in rule generation. This study illustrates that SmartLLMSentry significantly enhances the speed and accuracy of vulnerability detection through LLM-driven rule integration, offering a new approach to improving Blockchain security and addressing previously underexplored vulnerabilities in smart contracts.

*Keywords— Smart contract, Vulnerability, Software security, Blockchain, Large Language Models*

## I. INTRODUCTION

In the rapidly evolving digital ecosystem, smart contracts have emerged as a transformative technology, fundamentally altering how agreements are executed and enforced. Embedded within blockchain technology, these self-executing contracts automatically carry out the terms of an agreement once predetermined conditions are met. This automation not only streamlines processes but also significantly reduces the need for intermediaries, fostering a more direct and efficient transactional environment.

The burgeoning reliance on these technologies underscores the critical need for robust security measures. Smart contracts are not impervious to vulnerabilities; their open-source nature and the immutability of the blockchain make them susceptible to various types of attacks, which could lead to significant financial and reputational damage [1]. Common vulnerabilities include reentrancy attacks [2], transaction-ordering dependence [3], timestamp dependence [3], and several others [3] that can compromise the intended functionality and security of these applications [4].

The increasing complexity and deployment of smart contracts across various industries [5] have highlighted the necessity to automate the process of finding vulnerabilities. As these contracts become more intricate, manually identifying potential security risks becomes less feasible and more error-prone [6]. This recognition has spurred interest in the research community, leading to significant efforts toward developing automated tools that can efficiently and accurately detect vulnerabilities at scale. The drive for automation is not merely a matter of convenience but a critical requirement to maintain the integrity and trustworthiness of blockchain applications. As the adoption of smart contracts continues to grow, fueled by their potential to revolutionize traditional business models, the development of advanced automated vulnerability detection frameworks becomes imperative to ensure their safe operation.

In response to the need for more effective security measures, various techniques have been employed to detect vulnerabilities in smart contracts. Traditional methods primarily involve static analysis, which analyzes the contract's code without executing it. However, those frameworks operates by comparing code against a set of predefined rules that describe known vulnerabilities. However, this approach has inherent limitations, primarily because it relies heavily on the accurate definition of these rules by human experts. As a result, this technique inherently depends on human experts to continually define new rules for newly detected vulnerabilities. The reliance on expert input not only limits the speed at which new threats can be addressed but also underscores a fundamental constraint: the system's efficacy is tied to the timely and accurate update of its rule set. Without constant refinement and expansion of these rules, the framework risks missing novel vulnerabilities, highlighting the critical need for ongoing expert involvement to maintain its effectiveness.

In recent years, the capabilities of LLMs such as GPT have garnered significant interest due to their impressive computational intelligence. These models are developed by training on vast datasets and require substantial computational resources. With billions of parameters, LLMs can discern intricate patterns and demonstrate a level of general intelligence across a range of tasks previously considered challenging for AI to achieve in the short term. A notable instance of this technology is ChatGPT, a sophisticated chatbot powered by openAI [7]. It engages users in human-like interactions, providing responses across various domains. Given these advancements, recent research has explored the potential of using ChatGPT to enhance the detection of software vulnerabilities, suggesting a new application area where LLMs could significantly impact [8, 9].

Following the same logic and to solve rule based frameworks limitations, this paper launches a study on the performance of rule generation using an in-context learning version of ChatGPT with different prompt designs. gpt-4o-mini-2024-07-18 is the newest version of ChatGPT when we conduct this study [10].

The contributions of this work:

- We have developed an automated system for rule creation within, which eliminates the need for continual expert intervention and accelerates the integration of new rules. This innovation enhances the responsiveness of the system to emerging vulnerabilities and streamlines the update process.

- In our research, we have identified and analyzed three new common root causes for vulnerabilities which, to the best of our knowledge, have not previously been explored in the scientific literature. We provide a detailed examination of their root causes, contributing valuable insights into the underlying issues that lead to these security flaws.

- We have built a Dataset of Five different smart contract vulnerabilities that could be used by research to test or train their models.

- We have successfully created new rules into SmartLLMSentry Analyzer component that extend its capability to detect three additional types of vulnerabilities. This expansion not only improves the comprehensiveness of SmartLLMSentry's security measures but also increases its utility in safeguarding smart contracts.

- We conducted a comprehensive study on the effectiveness of various versions of GPT models in detecting vulnerabilities within smart contracts. This investigation helps in understanding which model iterations perform best in this specific context and guides future implementations of LLM technologies in security applications.

- We explored the impact of dataset size on the performance of GPT models specifically in the context of generating rules for vulnerability detection. Our findings reveal significant insights into how the quantity and quality of training data influence the accuracy and reliability of the rules produced.

- Finally, we provide a detailed analysis of the advantages and limitations of employing ChatGPT for detecting vulnerabilities in smart contracts. This analysis offers a balanced view, highlighting where ChatGPT excels and where it may require further tuning or supplementation to meet the demands of this application.

This paper is organized as follows: Section 2 provides a review of related work in both traditional software and smart contract security, including the application of LLMs in vulnerability detection. This section also outlines the five specific vulnerabilities targeted in our study and examines their common root causes. Section 3 details the data collection and preprocessing methods employed in the research. Following this, Section 4 describes the prompt engineering process, including the various prompts utilized during experimentation. Section 5 presents and discusses the research findings, while Section 6 addresses potential threats to validity. Finally, the paper concludes with a summary of the results and suggestions for future research.

## II. RELATED WORK

### A. Static analysis techniques

Due to both their immutability feature and the managed sensitive data, smart contract vulnerabilities should be checked and fixed before a production deployment. The rise in their complexity make manual vulnerability detection of smart contract not efficient and require to be completed by an automated scan [11] Therefore, multiple previous researches [12, 13, 14, 15, 16] have been performed to build frameworks that automatically discover those vulnerabilities in the development phase. Feist et al, [17] have built slither, a static smart contract analyzer that is capable of detecting vulnerabilities like Shadowing [3], Uninitialized variables [3], Reentrancy [18] and a variety of other known security issues, such as suicidal contracts, locked ether, or arbitrary sending of ether. Grech et al., built MadMax [19], a static analysis framework capable of detecting out-of-gaz related vulnerabilities [3]. Nguyen et al., in 2020 have built sFuzz [20] a dynamic analysis framework capable of detecting vulnerabilities like Reentrancy, Timestamp Dependency [3], Block Number Dependency [3], Integer Overflow [3], and others. Ren et al., [21] have also built a static vulnerability detection framework called Solidifier capable of detecting vulnerabilities like, Reentrancy, Timestamp Dependency, Front Running, Integer Overflow, and others.

However, a significant limitation of those frameworks lies in the manual creation of rules for vulnerability detection. This manual process can be time-consuming and prone to human error, limiting the framework's ability to adapt to new and evolving threats. In this paper, we propose SmartLLMSentry to solve those limitations using LLMs. By leveraging LLMs, we aim to automate the rule generation process, thereby improving the efficiency and accuracy in detecting a broader range of vulnerabilities. This approach not only addresses the current limitations of static analyzers but also enhances their adaptability to emerging security threats in smart contracts.

### B. Machine learning and LLMs

The emergence of machine learning and its proven success across various domains has prompted researchers to apply these techniques to the detection of smart contract vulnerabilities. One of those researches was a framework built by Zhipeng et al, called SMARTEMBED [22], which is a web service tool for Solidity developers. SMARTEMBED uses code embeddings and similarity checking to detect repetitive code and clone-related bugs. Applied to over 22,000 Solidity contracts, it identified a 90% code clone ratio and detected 194 clone-related bugs with 96% precision. Yu et al. developed DeeSCVHunter [23], a modular framework for detecting smart contract vulnerabilities, introducing the concept of Vulnerability Candidate Slicing (VCS) which enriches semantic and syntactic features to enhance deep learning model performance. However, DeeSCVHunter is limited to detecting re-entrency and timestamp vulnerabilities only. Wu et al. [24] introduced a smart contract representation method based on key data flow graph information to capture essential features for vulnerability detection while mitigating overfitting during training. They proposed a tool named Peculiar, which enhances detection performance by utilizing critical data flow graphs. However, the construction of these graphs is complex, and Peculiar's detection capabilities are limited to identifying reentry vulnerabilities in smart contracts. Another research performed by Nami et al, [25] resulted in building an effective framework against code rewriting attacks. Eth2vec is a machine learning-based static analysis tool, contrarely to the other models Eth2vec works with Ethereum Virtual Machine (EVM) bytecodes. Despite its strengths, Eth2vec requires manually crafted vulnerability features to maintain detection accuracy, which can be labor-intensive and limits its adaptability to new types of vulnerabilities.

Lately, AI has undergone a revolution thanks to LLMs, which have demonstrated extraordinary skill in a variety of tasks [26, 27, 28]. LLMs are founded on the principle of language modeling (LM), which involves modeling the generative likelihood of word sequences to predict future tokens. Unlike traditional LM approaches, LLMs are trained on vast datasets using powerful computational resources. Their versatility and adaptability are attributed to their billion-scale parameters [29], a level of complexity and scale unprecedented in previous models. This extensive parameterization allows LLMs to achieve remarkable performance across a wide range of language tasks. The most popular LLMs to the date of writing this paper are OpenAI's GPT [7, 30], Gemini [31] and Claude [32]. Some of them provide an API access to finetune their model, while others do not [33]. Many recent researchs have used the LLM technology to automate the process of vulnerability detection in softwares.

LLMs have been increasingly utilized in vulnerability detection. Sihao et al. [34] provide an in-depth analysis of using LLMs, such as GPT-4, to identify vulnerabilities in smart contracts, emphasizing the challenge of balancing accurate vulnerability detection with minimizing false positives. Their empirical research demonstrates that increased randomness in responses enhances the likelihood of correct detection but also raises the incidence of false positives. To mitigate this issue, the study introduces GPTLens, an adversarial framework that divides the detection process into two phases: generation and discrimination. In this framework, the LLM functions both as an auditor and a critic. This dual-role approach, designed to expand the scope of vulnerability detection while reducing inaccuracies, significantly outperforms traditional methods, establishing GPTLens as a versatile LLM-based solution that does not require specialized expertise in smart contracts. Additionally, LLM4Vuln [35] conducts a more detailed study aimed at decoupling LLMs' vulnerability reasoning capabilities from their other functionalities, yielding promising results in the application of these models for vulnerability assessment.

However, Static algorithm-based detectors offer significant advantages in terms of deterministic and consistent results, efficiency, and transparency. Unlike LLM-based analyzers, static detectors produce consistent outputs for the same input, ensuring reliability in critical environments. They are also more efficient, requiring fewer computational resources, making them faster and more cost-effective to operate.

Additionally, their decision-making process is transparent and interpretable, as they rely on predefined rules and patterns. This clarity allows developers to easily understand why a particular vulnerability was detected, simplifying the debugging process and enabling quicker remediation of identified issues. Despite all those advantages, static analyzers like any algorithm based technology suffer from a key limitation related to depending on expert knowledge to create detection rules [36]. Therefore, to avoid this dependency we use LLMs to generate the right algorithm to detect known vulnerabilities and enhance SmartLLMSentry framework detectors integration.

### C. Prompt Engineering for In-Context Learning

In-Context Learning (ICL) is a method by which a pre-trained model, like a LLM, can perform new tasks without needing additional training or fine-tuning [37]. Instead, the model is given a prompt that includes examples of the task it needs to perform [37]. The model then uses these examples to infer the task and generate appropriate responses [38]. This method leverages the model's existing knowledge and patterns learned during its training to adapt to new situations quickly [38]. For instance, if the model is given a few examples of a translation task within a prompt, it can learn to translate new sentences by understanding the context provided [38].

Fine-tuning a model, on the other hand, involves adjusting the model's parameters using a new dataset specific to a particular task [39]. This process requires additional training and computational resources [40, 41]. Fine-tuning typically allows for higher accuracy and better performance on the specific task but requires more data and time [40]. It also makes the model specialized, potentially reducing its generalization ability to other tasks [41].

We chose In-Context Learning over fine-tuning due to its flexibility and efficiency [38]. In-Context Learning allows us to leverage the model's pre-existing knowledge without the need for extensive retraining [38]. This is particularly useful when quick adaptation to new tasks is needed or when the task-specific data is limited. Additionally, ICL maintains the model's generalization capabilities, making it more versatile across various tasks. The ability to handle multiple tasks without requiring separate fine-tuned models simplifies deployment and reduces the computational overhead [41], making it a more practical choice in dynamic environments.

Prompting-based learning has emerged as a dominant paradigm in the utilization of language models. Rather than relying on objective engineering to adapt pre-trained LMs for downstream tasks, prompting-based learning reconfigures these tasks using a textual prompt, aligning them more closely with the tasks the LM was initially trained to solve [42]. Research has demonstrated that well-structured prompts can significantly enhance the performance of LLMs across a variety of downstream tasks [43, 44]. Consequently, a diverse array of prompt design strategies has been developed to further optimize the effectiveness of this approach [42].

Concerning prompt formulation, certain studies have focused on exploring the search for optimal discrete prompts [45, 46, 47]. However, other researchs have used continuous vector as prompts [48, 49, 50]. Several studies have investigated the impact of prompts on generative models. For instance, Liu et al. [51] examine how different prompts influence the generation of visualizations by LLMs [52]. Additionally, Liu et al. [53] propose various prompt designs tailored for two distinct code generation tasks. Recent research has also begun to explore the potential of ChatGPT in the context of software vulnerability detection. Cao et al. [8] design enhanced prompt templates to leverage ChatGPT for deep learning-based program repair. White et al. [54] explore various prompt patterns aimed at improving requirements elicitation, rapid prototyping, code quality, deployment, and testing. However, to our best knowledge none of the research have analyzed the performance of combining static analysis and LLMs power to detect smart contract vulnerabilities.

*D. Smart contract vulnerabilities*

Effective detection of smart contract vulnerabilities relies on understanding their common root causes. This section explores the frequent origins of five vulnerabilities that will be used as a training data or testing data. By examining these underlying issues, we aim to enhance the application of LLM prompt engineering in identifying vulnerabilities. Recognizing these root causes will not only improve the accuracy of our LLM-based detection methods but also contribute to the development of more robust and secure smart contracts.

In our research, we opted for a random selection of targeted vulnerabilities to ensure that our enhanced framework is tested against a diverse and representative set of potential threats. The primary objective was to avoid any bias that might arise from deliberately selecting certain types of vulnerabilities, which could skew the results and limit the generalizability of our findings. By using a random selection process, we aimed to simulate real-world conditions where vulnerabilities can vary widely in nature and impact. This approach allows us to assess the robustness and adaptability of our framework in generating and integrating rules of a broad spectrum of security issues, ultimately leading to more comprehensive and reliable improvements. Moreover, randomness in selection helps in avoiding overfitting our framework to a specific subset of vulnerabilities, ensuring that the LLM ICL is applicable across various scenarios rather than being tailored to specific cases.

*1) Array length manipulation (SWE-161)*

In older versions of Solidity, it was possible to manipulate the "length" field of an array using standard arithmetic operations, leading to serious vulnerabilities. Specifically, the "length" field, which determines the size of the array, could be decremented using code such as anArray.length--;. If this operation caused the length to underflow, it would result in the array size being set to the maximum possible integer value [55].

This underflow vulnerability could have severe consequences, potentially disabling a smart contract by allowing unauthorized or unintended access to memory locations far beyond the array's intended bounds. Such an issue could be exploited to disrupt the contract's functionality or even lead to the loss of critical data, as the contract might behave unpredictably [3].

*2) Message call with hardcoded gas amount (SWE-134)*

Hard forks in blockchain networks can lead to significant changes in the gas costs associated with executing Ethereum Virtual Machine (EVM) instructions. These changes can disrupt existing smart contracts that were deployed with the assumption of stable gas prices. As a result, hardcoding a fixed gas amount for specific contract operations can become a critical vulnerability over time. If the gas cost of certain EVM instructions increases due to a hard fork, the previously sufficient hardcoded gas limits may no longer cover the necessary execution, leading to a DoS condition [4].

This issue is particularly problematic because smart contracts, once deployed, are immutable and cannot be easily updated to accommodate new gas costs. For instance, the implementation of EIP-1884, which increased the gas cost of the SLOAD instruction [56], inadvertently disrupted the functioning of many existing smart contracts [57]. These contracts had hardcoded gas values based on the previous cost of SLOAD, and the increase rendered them unable to execute certain operations, leading to failures and potential vulnerabilities. The potential for such disruptions underscores the risks associated with hardcoding gas values in smart contracts. As blockchain protocols evolve and undergo hard forks, the assumptions underlying gas costs may no longer hold, and contracts that depend on these assumptions can become vulnerable to DoS attacks [3].

*3) Transaction order dependence (SWE-114)*

In blockchain networks where the validation order of transactions is not strictly enforced, nodes often prioritize transactions with higher fees to optimize their financial returns. This practice introduces a potential security risk, particularly for smart contracts that depend on the specific sequence in which transactions are validated. When the correct functioning of a smart contract is contingent on the order of transactions, a race condition may arise, leading to what is known in the blockchain domain as a Transaction Order Dependence (TOD) vulnerability [3].

A TOD vulnerability occurs when the outcome of a smart contract can be manipulated by altering the order of transactions [58]. This type of vulnerability is particularly relevant in scenarios where multiple transactions interact with the same contract in a short time frame. An attacker can exploit this by submitting a transaction with a higher fee to ensure it is processed before others, thereby gaining an unfair advantage or causing unintended consequences within the contract's logic [3, 58].

A common example of this vulnerability is found in the use of the approve() function in ERC-20 tokens. The approve() function allows a token holder to grant another address permission to spend a specified amount of their tokens [59]. However, if two transactions to modify the approved amount are processed out of order, it can lead to inconsistencies in the allowance, potentially enabling unauthorized spending or other malicious activities. This example illustrates how TOD vulnerabilities can compromise the security and expected behavior of smart contracts, emphasizing the importance of careful design and implementation to mitigate such risks.

*4) Locked Money (SWE-138)*

Smart contracts that accept tokens or funds without incorporating a mechanism for their retrieval are susceptible to vulnerabilities that can lead to the permanent locking of assets. This vulnerability is inherent in any contract that receives tokens or funds but lacks a clear procedure for their return or withdrawal. Such a design flaw can result in funds being trapped indefinitely, as the contract does not provide a way to reverse or access the deposited assets [3].

Furthermore, the use of the _mint() function to create and allocate tokens to unknown addresses introduces additional risks. If the recipient address does not support the specific type of token being minted, the tokens may become inaccessible, resulting in a loss of resources [59]. This issue arises because the receiving address may lack the necessary functionality or compatibility to handle the minted tokens effectively [59].

To address this problem, it is advisable to use a safeMint function instead. The safeMint function incorporates additional safety checks to ensure that the recipient address can handle the token being minted [59, 60].

*5) Improper handling of exceptions (SWE-140)*

One of the common root causes of the SWE-140 vulnerability is the use of the send or transfer functions for sending funds in smart contracts. These functions automatically impose a limit of 2300 gas for their execution [61], which acts as a safeguard against reentrancy attacks by restricting the amount of code that can be executed in the fallback function [3]. However, with the enhanced protections now available in modern smart contracts, it is increasingly considered safe to use the call function instead of send or transfer. The call function does not impose the same gas limits, allowing the receiving contract to execute more complex code within its fallback function [62, 63].

## III. METHOD

### A. Framework design

The architecture of SmartLLMSentry is characterized by 2 main parts Figure 1. The first one is the framework core which is also built of three interdependent components as shown in Figure 1. The pre-compiler, compiler, and analyzer, each contributing to a holistic and thorough smart contract analysis pipeline. The pre-compiler module, positioned as the inaugural stage, serves as the entry point for the analysis process. It accommodates diverse inputs, allowing users to specify either a GitHub URL or a smart contract address on the Ethereum Blockchain. The flexibility in input sources caters to various development and deployment scenarios.

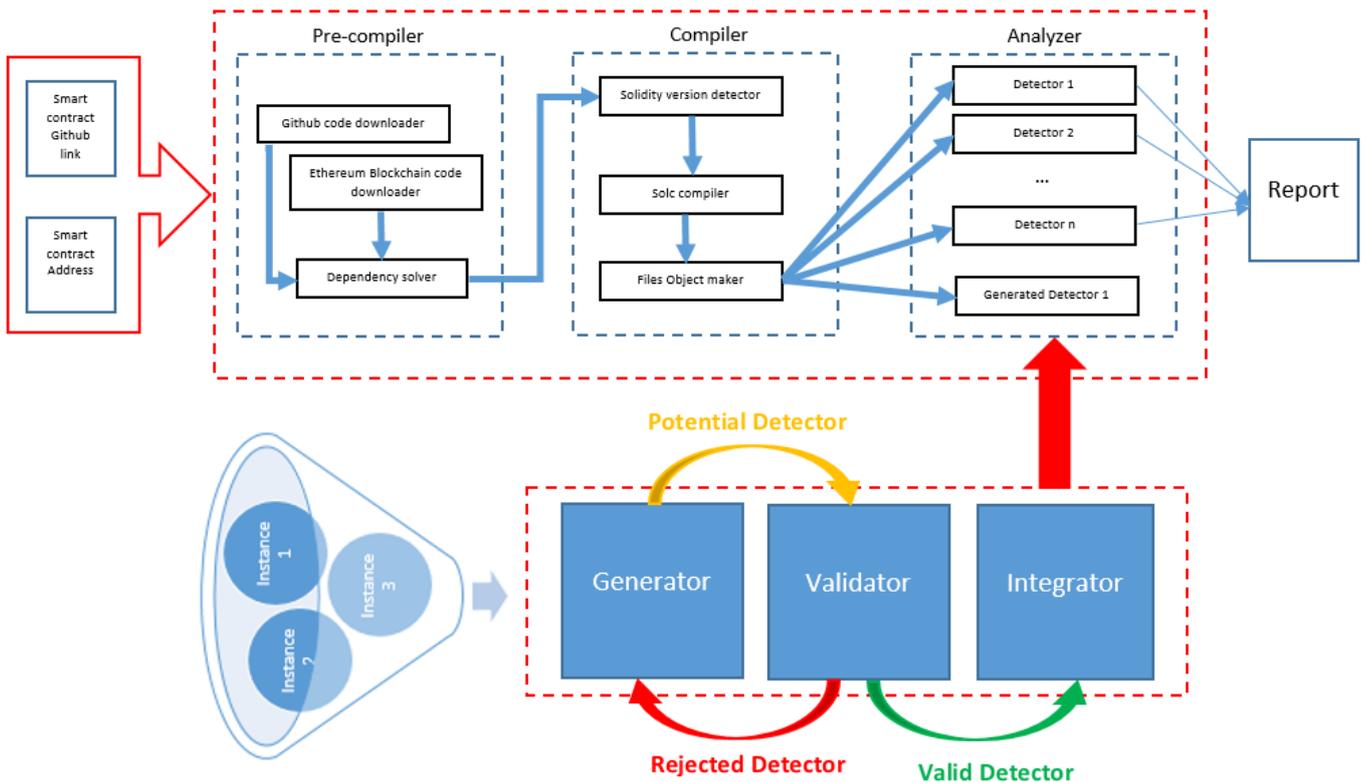

**Figure 1.** SmartLLMSentry Framework Design

Upon receiving the input Figure 1, the pre-compiler diligently undertakes the task of retrieving the specified codebase. This not only includes the primary smart contract but also extends to encompass all associated libraries, packages, and relevant code dependencies. The comprehensive inclusion of these elements ensures a self-contained and cohesive environment for subsequent compilation stages. The goal is to preemptively address any potential compilation challenges arising from dependencies, thereby enhancing the efficiency of the overall analysis process.

Moving forward, the compiler Figure 1, as the core processing unit, takes charge of the compilation process. Beyond the conventional compilation task, it plays a pivotal role in generating crucial artifacts that form the basis of subsequent analyses. Specifically, the compiler extracts both the Abstract Syntax Tree (AST) and the Control Flow Graph (CFG) from the compiled code. These representations provide a high-level abstraction of the smart contract's structure and the sequence of control flow, offering valuable insights into the code's intricacies.

The extracted AST and CFG, now enriched with semantic information, are then seamlessly transmitted to the static analyzer Figure 1. This component constitutes the heart of SmartSentry, employing static analysis techniques such as AST and control flow analysis and dataflow analysis. The static analyzer meticulously examines the received artifacts, conducting an in-depth analysis to detect potential security vulnerabilities. It scrutinizes the code for patterns indicative of common vulnerabilities, potential exploits, and deviations from best coding practices. The analyzer has been systematically designed with a focus on modularity, adhering to a structured approach that isolates each vulnerability's detection mechanism from others within SmartSentry. This deliberate separation of detectors ensures a modular and flexible architecture, facilitating seamless integration and scalability for future enhancements.

The segmentation of vulnerability detectors Figure 1 allows for targeted improvements or the addition of new detectors without necessitating extensive modifications to the existing framework. This design rationale is founded on the principle of providing an adaptable framework where individuals can effortlessly construct detectors tailored to specific vulnerabilities. Leveraging the built-in functionalities for AST analysis or dataflow analysis, stakeholders have the capability to construct specialized detectors in accordance with the unique requirements of distinct vulnerabilities. This modular design philosophy not only enhances the extensibility and versatility of the framework but also fosters a collaborative environment conducive to continual advancements in smart contract security analysis.

The second part is responsible for a continues and automated enhancement of the framework and it is also built upon 3 other components connected directly to the Analyzer Figure 1. Once a new vulnerability is detected in the wild, a number of vulnerable code instances are sent to the Generator. This component is responsible for creating detectors for specific vulnerabilities. It receives a set of instances representing a particular vulnerability and utilizes these instances to generate a detector tailored to identify that vulnerability. The generator is essential for producing initial candidate detectors based on real-world data, ensuring that the detectors are relevant and applicable to the specific vulnerabilities being addressed.

Following the generation of detectors, the validator assesses their performance by evaluating their accuracy using the same instances provided to the generator Figure 1. The validator is critical for ensuring the reliability of the detectors. If a detector achieves an accuracy rate below 80%, it is deemed inadequate and rejected. The generator is then tasked with producing a new detector to meet the accuracy threshold. This iterative validation process is vital for maintaining high-quality detection capabilities and preventing false positives or negatives.

Detectors that surpass the 80% accuracy threshold are accepted by the validator and subsequently passed to the integrator Figure 1. The integrator's role is to incorporate these validated detectors into the SmartSentry framework. This component ensures that new, high-performing detectors are seamlessly integrated into the existing system, enhancing the overall capability of SmartSentry. To facilitate debugging and maintenance, the integrated detectors are labeled with the term "generated," indicating their origin and allowing for easier identification and troubleshooting of any issues.

*B. Prompt Design*

This section outlines the prompts we have designed to improve ChatGPT's performance in detecting software vulnerabilities. For clarity, we denote each prompt as $Px$, where $x$ represents the specific components of the prompt. Each component $x$ will be detailed as we introduce and explain the corresponding prompt design for the first time.

TABLE I. LIST OF USED TRAINING PROMPTS

| Code | Prompt |
|---|---|
| $P_b$ | Write the if condition to detect this instruction. |
| $P_{rb}$ | You are a smart contract security auditor. write the if condition to detect this instruction. |
| $P_{rcb}$ | You are a smart contract security auditor. Using Solidity-ast and typescript, write the if condition to detect this instruction.<br>The output should only contain the if condition. |
| $P_{rcbi}$ | You are a smart contract security auditor. Using Solidity-ast and typescript, write the if condition to detect this instruction.<br>The output should only contain the if condition.<br>Keep in mind that The following ast structures could have different values depending on their nodeType:<br>Expression.nodeType<br>rightExpression.nodeType<br>leftExpression.nodeType<br>leftHandSide.nodeType<br>rightHandSide.nodeType |

*1) Basic Prompting*

Firstly, to conduct vulnerability detection via ChatGPT, it is essential to have a basic prompt (Pb). We use the following basic prompt in this study, and we ask ChatGPT to generate the require if condition that should be injected in SmartLLMSentry Analyzer component to be able to detect this vulnerability in any other source code.

Following OpenAI's gpt-best-practices document [10], we further propose the role-based basic prompt Prb to remind ChatGPT of its job (i.e., smart contract vulnerability detection) so that it focuses on the vulnerability issue:

As the generated if condition will be included into our previously developed framework, we needed to give more context information to gpt model to further optimize its output and focus only on the if condition.

*2) Prompting with Auxiliary Information*

In Solidity, the AST structure of multiple nodes can vary significantly based on the nodeType. Through extensive testing, we observed that ChatGPT does not consistently adhere to the correct Solidity AST structure. Therefore, we decided to incorporate additional information regarding the Solidity AST structure into our prompt. Specifically, we augmented the Prompt with auxiliary information about the AST structure of Solidity source code to improve its accuracy and reliability in generating and interpreting Solidity code.

## IV. Dataset

To train our model and apply the prompts we developed, we required a suitable dataset. However, given that smart contract security is a relatively new field, and the vulnerabilities we selected are largely unexplored, we initially lacked the necessary data to train our model effectively. In this section, we explain the various steps involved in building the dataset and how we addressed the challenges encountered during the process. This includes our strategies for data collection, preprocessing, and augmentation, which were essential to overcoming the limitations of data scarcity in this emerging field.

### A. Data Collection

The first step in developing our dataset involved collecting initial data from multiple sources. We gathered raw data from two primary repositories: GitHub and live smart contracts available on Etherscan. GitHub provided a valuable source of open-source smart contracts, while Etherscan offered insights into deployed contracts on the Ethereum network. This combination of sources allowed us to capture a diverse range of smart contracts and associated vulnerabilities.

To enhance and diversify our dataset, we implemented an additional strategy of injecting vulnerable code snippets into forged functions. By incorporating known vulnerabilities into these synthetic functions, we were able to artificially amplify the dataset and simulate various scenarios of interest. This approach not only increased the volume of data but also ensured that our dataset covered a broader spectrum of potential vulnerabilities, thereby enriching the training and testing phases of our model.

The dual approach of sourcing real-world data and augmenting it with synthetic examples enabled us to build a more comprehensive and robust dataset, addressing the challenges posed by the initial data scarcity in the field of smart contract security.

### B. Data Pre-processing

In this study, we collected and utilized a dataset to train and evaluate our framework for detecting smart contract vulnerabilities through LLM prompt engineering. The data collection and preparation involved several critical steps to ensure the quality and effectiveness of our model.

First, we removed duplicate code snippets from the dataset to maintain its uniqueness and relevance. Given the constraints on input length imposed by ChatGPT, we carefully managed the size and format of our data. The dataset initially comprised 150 instances, which we divided into two subsets: 112 instances for training and 38 instances for testing. This partition allowed us to build and refine our model while reserving a separate set of data for rigorous evaluation.

Based on OpenAI's guidelines, which recommend a range of 50 to 100 instances for effective ICL, we structured our training data accordingly. We created two training groups: the first group included 100 instances, used to train the model initially. The second group comprised these 100 instances plus an additional 12 instances, totaling 112 examples, to investigate whether increasing the dataset size would improve the model's performance. The data was formatted into JSON Lines (JSONL) to ensure compatibility with ChatGPT's input requirements. This format facilitated efficient data processing and integration with the model. By using this structured approach, we trained our model on the prepared data and evaluated its performance using the reserved test set.

## V. Experimentation And Results Descussion

In this section, we report and analyze the experimental results inorder to answer the following research questions (RQ):

RQ1: Can ChatGPT generate valid detection code for specific vulnerabilities?

RQ2: How the amount of training data could affect the effectiveness of the model?

RQ3: Which version of chatgpt is best suited for smart contract vulnerability detection rule generation?

### A. Experimental Settings

*1) Finetunning Parameters*

The ChatGPT fine-tuning platform provides several critical parameters that can significantly influence the model's final performance. Understanding and carefully configuring these parameters is essential for optimizing the model for specific tasks. The first parameter, Seed, is crucial for ensuring the reproducibility of the fine-tuning process. By setting a specific seed value, researchers can control the randomization involved in training, allowing for consistent results across different runs. The following tables present the different seeds used during our experimentation to reproduce the same exact results.

TABLE II. Seeds Used In Each Training For Dataset Size Of 100.

| | Dataset size 100 | | | |
|---|---|---|---|---|
| | **Pb** | **Prb** | **Prcb** | **Prcbi** |
| **gpt3** | 184181018 | 1229186277 | 2050353472 | 1082851968 |
| **gpt4** | 1546849632 | 2143196420 | 1653381796 | 542758792 |

TABLE I. Seeds Used in Each Training for Dataset Size Of 112

| | Dataset size 112 | | | |
|---|---|---|---|---|
| | **Pb** | **Prb** | **Prcb** | **Prcbi** |
| **gpt3** | 317425969 | 1011665401 | 1553307650 | 294133873 |
| **gpt4** | 1692427035 | 1804209602 | 1109655604 | 956977286 |

The second parameter, **Batch Size**, determines the number of training examples used in one iteration of the model's learning process. A smaller batch size, enables the model to update more frequently, which can be beneficial for tasks requiring fine-grained adjustments. However, it also demands more computational resources and time. The third parameter, **Learning Rate Multiplier**, is a scaling factor applied to the base learning rate. This multiplier adjusts the speed at which the model learns during training. A higher learning rate multiplier, allows the model to converge more quickly, but it also risks overshooting the optimal solution, leading to suboptimal performance. Lastly, the **Number of Epochs** represents the total number of times the model will cycle through the entire training dataset. Setting the number of epochs to higher value indicates that the model will undergo multiple training cycles, helping it generalize better by repeatedly learning from the dataset.

Initial experiments indicate that the optimal configuration for our use case includes setting the number of epochs to 3, the batch size to 1, and the learning rate multiplier to 2. This configuration has shown to produce the best results in terms of model performance and accuracy in our specific application.

*2) Used GPT versions*

In our study, we have compared ChatGPT models effectiveness:

- gpt-3.5-turbo-1106
- gpt-4o-mini-2024-07-18

The choice of gpt models comes from the a scientific research performed by Wenpin et al. [64] that has clearly shown that Chatgpt outperform all the existing LLM models both opensource and closed source solutions in code generation.

*3) Evaluation Metrics*

To evaluate the effectiveness of our LLM models, we will use the Exact Match (EM), which is a commonly used metric to evaluate the effectiveness of an LLM model in generating valid code. However, we define the Exact match of two code snippet as the situation where both code has the exact same logic or the exact same syntax.

$$\text{Exact Match (EM)} = \frac{\text{Number of Exact Matches}}{\text{Total Number of Examples}} \times 100$$

**Number of Exact Matches:** Count the number of generated code snippets that exactly match the expected output code snippets.

**Total Number of Examples:** The total number of code snippets evaluated.

The total number of examples in our experimentation test inputs is **38**.

Here are the results of our experimentations:

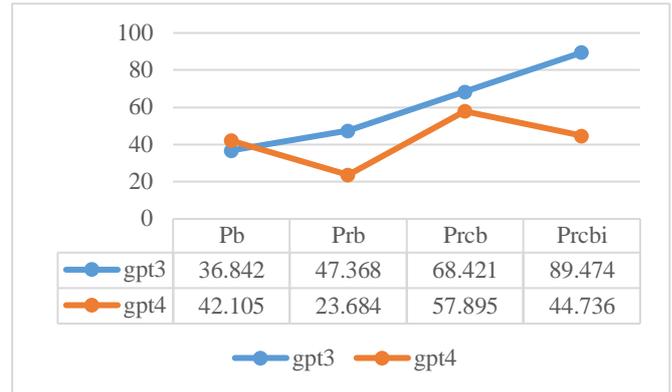

**Figure 1.** EM of both trained models with different Prompts and 100 dataset inputs.

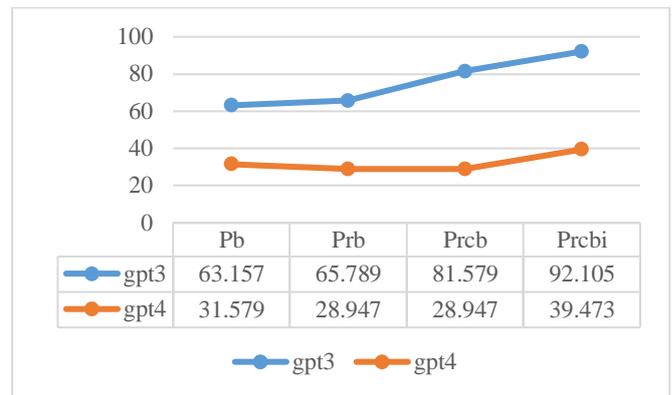

**Figure 2.** EM of both trained models with different Prompts and 112 dataset inputs.

The results presented in Figures 2 and 3 illustrate the performance of GPT-3 and GPT-4 in terms of EM accuracy across different dataset sizes and prompt types. In Figure 2, which evaluates the models with a dataset of 100 instances, GPT-3 achieved an EM score of 36.8% for prompt type Pb, 47.4% for Prb, 68.4% for Prcb, and 89.5% for Prcbi. In contrast, GPT-4's EM scores were lower, with 42.1% for Pb, 23.7% for Prb, 57.9% for Prcb, and 44.7% for Prcbi.

Figure 3 shows the results for a larger dataset of 112 instances. GPT-3's performance improved significantly, achieving EM scores of 63.2% for Pb, 65.8% for Prb, 81.6% for Prcb, and 92.1% for Prcbi. Conversely, GPT-4's performance was notably poorer, with EM scores of 31.6% for Pb, 28.9% for Prb, 28.9% for Prcb, and 39.5% for Prcbi.

*B. Effectiveness of finetunned GPT models (RQ1)*

The data reveal that GPT-3 consistently outperforms GPT-4 across all prompt types and dataset sizes. Even with additional training data, GPT-4's performance did not improve and, in fact, declined, suggesting that simply increasing the dataset size did not enhance its accuracy. Notably, GPT-4 did outperform GPT-3 with a dataset of 100 instances and a basic prompt. Nevertheless, under the specific experimental conditions and dataset configurations employed, GPT-3 exhibited superior exact match accuracy compared to GPT-4. This indicates that GPT-3 is more effective at leveraging available data to produce accurate results, while GPT-4 may need further refinement or adjustment to improve its performance.

C.     *The training data volum impact (RQ2)*

The impact of increasing the training dataset on both GPT-3 and GPT-4 models is evident from the data provided. With a dataset of 100 instances, GPT-4 achieved higher EM accuracy compared to GPT-3 for a basic prompt. However, when the dataset size was increased to 112 instances, GPT-3's EM scores improved significantly across all prompt types and reached 92,1% EM, demonstrating its ability to effectively utilize more data for enhanced accuracy. In contrast, GPT-4's performance deteriorated with the larger dataset, as evidenced by lower EM scores across all prompt types. This decline suggests that simply increasing the dataset size did not benefit GPT-4's performance and may have revealed or exacerbated existing limitations within the model. These findings indicate that while GPT-3 showed a positive correlation between dataset size and performance, GPT-4's effectiveness was negatively impacted by the additional data.

D.     *The most suitable version? (RQ3)*

Based on the results obtained from our experimentation, GPT-3 is the most suitable model for our use case, particularly when additional data is available during the training phase. The data demonstrate that GPT-3 consistently outperforms GPT-4 in terms of EM accuracy across various prompt types and dataset sizes. With a larger dataset, GPT-3's performance improved significantly, showcasing its ability to leverage additional data effectively to enhance accuracy. In contrast, GPT-4's performance declined with increased dataset size, indicating that it may have limitations in handling larger datasets or may require further optimization to achieve comparable results.

Given these observations, GPT-3's superior performance and its positive response to increased data make it a better fit for scenarios where extensive training data can be utilized. GPT-4, while initially promising, did not demonstrate the same level of robustness or improvement with additional data, making GPT-3 the preferred choice for our specific use case.

## VI.   THREATS OF VALIDITY

A.     *Version of ChatGPT*

The results presented in this paper are based on experiments conducted with two specific versions of ChatGPT. Given that ChatGPT is an actively evolving platform, it is important to acknowledge that the findings and conclusions drawn here are limited to the performance characteristics of these particular versions. As OpenAI continues to update and refine ChatGPT, subsequent versions may exhibit different behaviors, capabilities, or performance metrics. Therefore, the conclusions drawn from this study might not be applicable to future iterations of the model or to older versions that were not assessed in our experiments.

Updates to the model may include improvements in architecture, training techniques, or data handling, which could significantly alter performance outcomes. Consequently, findings relevant to the versions tested in this study may become outdated or inaccurate as newer versions are released. Similarly, insights derived from the tested versions might not fully account for changes that could affect the model's effectiveness in different contexts or applications.

B.     *Different Solidity parser library*

Another significant threat to the validity of this study arises from the reliance on a specific Solidity parser library for analyzing smart contracts. The choice of parser can substantially influence the accuracy and completeness of the contract analysis due to variations in the implementation and features of different parser libraries.

In this study, we used a particular Solidity parser library called "solidity-ast-0.4.52" to process and analyze the smart contracts. However, there are several other available Solidity parser libraries, each with its own methodologies for parsing and interpreting Solidity code. These libraries may differ in their handling of syntax, support for various Solidity versions, and the extent of features they offer. Consequently, the results obtained using one parser might not be directly comparable to those obtained with another.

C.     *Vulnerability Types*

A notable threat to the validity of our results is the limited focus on only five specific types of vulnerabilities, chosen randomly. While these types provided valuable insights into ChatGPT's performance, the model's effectiveness might vary with different vulnerabilities. Given the diverse nature of smart contract vulnerabilities, including reentrancy attacks, integer overflows, and other issues, ChatGPT's performance could differ significantly when applied to other types not covered in this study. The model may show better results with vulnerabilities that align more closely with its training data or prompt design. Conversely, novel or complex vulnerabilities could present challenges not captured by our selected types. Therefore, while our findings are informative, future research should explore ChatGPT's performance with a broader range of vulnerabilities to assess its robustness and adaptability across different types.

## VII.   CONCLUSION

LLMs with advanced capabilities have made substantial impacts across various domains. This paper investigates the effectiveness of prompt-enhanced ChatGPT for detecting vulnerabilities in smart contracts, a crucial aspect for ensuring blockchain security and fostering trust. We developed several specialized prompts for vulnerability detection, incorporating additional contextual information, and focused on five types of vulnerabilities that, to our knowledge, had not been previously explored in the scientific literature. These vulnerabilities were selected randomly to minimize bias, and a dataset was constructed for both training and testing the model.

Our findings indicate that the trained model achieves a high accuracy rate of 91.1% in exact match scenarios when provided with sufficient information. However, the results also reveal that GPT-4 is less effective for our specific use case compared to GPT-3, with GPT-4 showing reduced accuracy in generating detection rules, even with an increased dataset.

While the results are promising and demonstrate the potential of LLMs and in-context training for vulnerability detection, further improvements are necessary. The current model's generated rules are not fully automated in the framework and still require some expert intervention. Future research will aim to explore additional types of vulnerabilities to further evaluate and enhance the model's effectiveness in diverse contexts.

ACKNOWLEDGEMENT

None.

FUNDING

This research did not receive any outside funding or support. The authors report no involvement in the research by the sponsor that could have influenced the outcome of this work.

AUTHORS` CONTRIBUTIONS

All authors have participated in drafting the manuscript. All authors read and approved the final version of the manuscript. All authors contributed equally to the manuscript and read and approved the final version of the manuscript.

CONFLICT OF INTEREST

The authors certify that there is no conflict of interest with any financial organization regarding the material discussed in the manuscript.

DATA AVAILABILITY

The data supporting the findings of this study are available upon request from the authors.